\documentclass[12pt]{iopart}

\usepackage{graphicx}

\begin{document}

\title{Ion-Photonic Frequency Qubit Correlations for Quantum Networks}

\author{Steven C. Connell$^1$, Jordan Scarabel$^1$, Elizabeth M. Bridge$^1$, Kenji Shimizu$^1$, Valdis Bl\=ums$^1$, Mojtaba Ghadimi$^1$, Mirko Lobino$^{1,2}$, Erik W. Streed$^{1,3}$}

\address{$^1$Centre for Quantum Dynamics, Griffith University, Brisbane, QLD 4111 Australia}

\address{$^2$Queensland Micro and Nanotechnology Centre, Griffith University, Brisbane, QLD 4111 Australia}
\address{$^3$Institute for Glycomics, Griffith University, Gold Coast, QLD 4222 Australia}
\ead{e.streed@griffith.edu.au}
\vspace{10pt}
\begin{indented}
\item[]\today
\end{indented}

\begin{abstract}
Efficiently scaling quantum networks to long ranges requires local processing nodes to perform basic computation and communication tasks. Trapped ions have demonstrated all the properties required for the construction of such a node, storing quantum information for up to 12 minutes, implementing deterministic high fidelity logic operations on one and two qubits, and ion-photon coupling. While most ions suitable for quantum computing emit photons in visible to near ultraviolet (UV) frequency ranges poorly suited to long-distance fibre optical based networking, recent experiments in frequency conversion provide a technological solution by shifting the photons to frequencies in the telecom band with lower attenuation for fused silica fibres. Encoding qubits in frequency rather than polarization makes them more robust against decoherence from thermal or mechanical noise due to the conservation of energy. To date, ion-photonic frequency qubit entanglement has not been directly shown. Here we demonstrate a frequency encoding ion-photon entanglement protocol in $^{171}$Yb$^+$ with correlations equivalent to 92.4(8)\% fidelity using a purpose-built UV hyperfine spectrometer. The same robustness against decoherence precludes our passive optical setup from rotating photonic qubits to unconditionally demonstrate entanglement, however it is sufficient to allow us to benchmark the quality of ion-UV photon correlations prior to frequency conversion to the telecom band.
\end{abstract}

%
\vspace{2pc}

\submitto{\NJP}
%
%
%

\section{Background}
Quantum networks \cite{Kimble_2008} are the key infrastructure for the long-range distribution of quantum information through the sharing of entanglement between distant locations. Such a network would enable a range of applications including quantum key distribution \cite{Bunandar_2018}, distributed quantum computing \cite{Cuomo_2020}, and blind quantum computing \cite{Fitzsimons_2017}. The main challenge preventing the sharing of quantum information over long distances is transmission losses. For a quantum state, these losses cannot be compensated via amplification as in classical communication networks but their long-range effects can be mitigated using more complex protocols.

The quantum repeater architecture \cite{Sangouard_2011} is a communications protocol which achieves long-distance communication by dividing the communication channel into shorter, lower-loss segments connected by nodes capable of storing quantum information and performing entanglement swapping \cite{Yuan_2008}. Satellite based quantum  communication provides a contrasting single segment architecture which has also been demonstrated over distances $>$1000 km \cite{Yin_2020,Yin_2017}, however requires clear lines of sight.

Trapped ions have implemented all the key functionalities of a quantum network node including local quantum information processing \cite{Wang_2017,Ballance_2016,Moehring_2007}, coupling of ion fluorescence into single mode fibres \cite{Maiwald_2012,Ghadimi_2017}, remote entanglement \cite{Maunz_2007}, and frequency conversion of single photons to the infrared telecom band \cite{Bock_2018,Kasture_2016,Krutyanskiy_2017,Krutyanskiy_2019}. At the local node trapped ions can store quantum states for $>$10 minutes \cite{Wang_2017}, realise high fidelity one and two qubit gates \cite{Ballance_2016}, and achieve deterministic ion-ion entanglement \cite{Moehring_2007}. High single ion fluorescence collection efficiency \cite{Maiwald_2012} and single mode ion-fibre coupling using a surface chip trap with integrated optics \cite{Ghadimi_2017} have also been demonstrated. Optical frequency conversion in nonlinear crystals has been used to interface the fixed fluorescence frequencies from ion qubits in the visible and ultraviolet with photonic qubits in the infrared telecom spectral region \cite{Bock_2018,Kasture_2016,Krutyanskiy_2017,Krutyanskiy_2019} where fused silica fibres exhibit extremely low losses.

Independent of the choice of hardware used in the network node, photons are the best candidate to be the courier for the transmission of quantum information. It is likely that these photons will be transmitted through optical fibres similar to the ones currently used for conventional optical telecommunications. While polarization is the most common type of encoding basis used for experiments with photonic qubits, real-world optical fibre networks has shown that this degree of freedom is substantially more susceptible to decoherence from environmental noise sources including vibrations, temperature fluctuations, and stress birefringence. These all would require active stabilisation \cite{Sun_2016} for transmission over long distances, with a substantial increase in complexity and potential fundamental noised based limits on fidelity. 

A more robust encoding basis for the photonic qubit is frequency. By mapping the qubit state directly on the energy of the photon, entanglement swapping can be performed on a nonpolarizing beam splitter without the need for active stabilization. However, manipulation of a photonic frequency qubit requires the use of active optical components, such as amplitude or phase modulators, instead of passive optical components including waveplates and polarizing beam splitters \cite{Lu_2019, Clemmen_2016}. 

Ion-ion entanglement mediated by frequency photonic qubit \cite{Moehring_2007}, and ion-photonic polarization qubit \cite{Blinov_2004,Crocker_2019} have been shown, but no direct measurement of ion-photonic frequency qubit has yet been measured. Characterizing the performance of the ion-frequency qubit state is essential for benchmarking a quantum network with arbitrary topology that uses this type of encoding and has applications in memory-assisted quantum communication \cite{Bhaskar_2020}.


Here we implement a protocol for the generation of ion-photon frequency entanglement and measure the correlations between ground state hyperfine levels of a single $^{171}$Yb$^+$ ion and the energy state of the emitted UV photon. For this purpose, we built a UV spectrometer with a maximum theoretical resolution of 1.73 GHz (FWHM), capable of clearly resolving the 12.6 GHz separation of the photonic frequency states. While entanglement could not be verified in our set-up, since it cannot implement arbitrary single qubit gate operations on the frequency degree of freedom of the photon, we demonstrated ion-photon correlations with a fidelity of 92.4(8)\%.

\section{Ion-photon entanglement protocol}
In a proposed architecture, trapped ion quantum networks \cite{Sangouard_2009} will link distant quantum nodes via optical fibres carrying photonic qubits. The first stage of entanglement swapping can be implemented using only photons and probabilistic Bell measurements. Subsequently, each quantum node, hosting a number of ions dependent on the topology of the network, will perform deterministic single qubit gates, entanglement swapping, and error correction. Telecom band photons can be interfaced with trapped ions using frequency conversion in nonlinear crystals \cite{Bock_2018,Kasture_2016,Krutyanskiy_2017,Krutyanskiy_2019}.

\begin{figure}[ht!]
    \centering
    \includegraphics{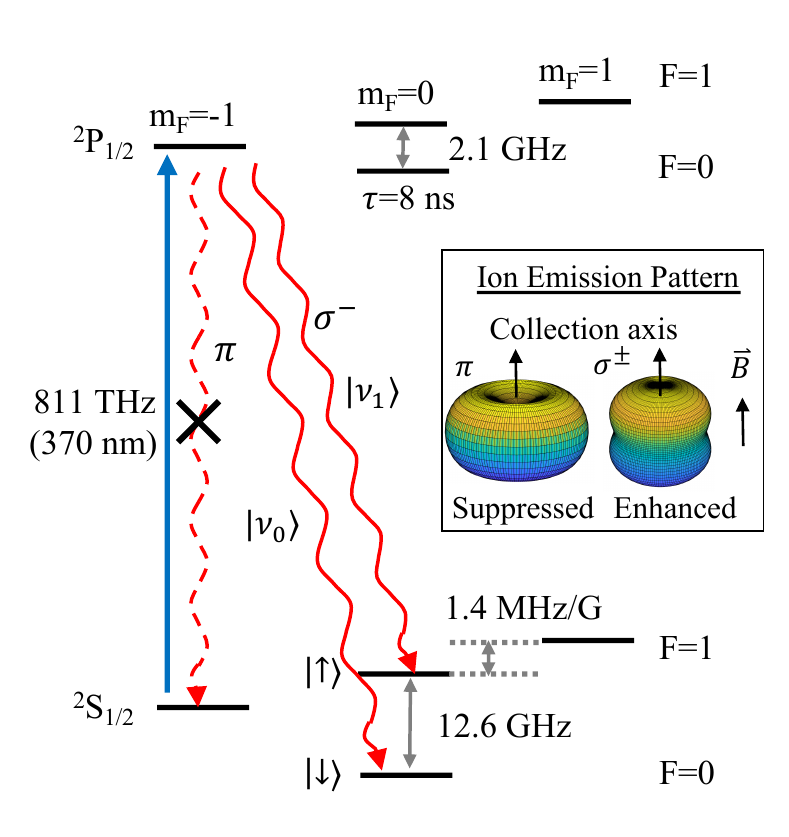}
    \caption{Relevant energy structure and transitions in $^{171}$Yb$^+$ which indicate the qubit states for the ion, the frequency qubit of the photon, and the optical pulse used to excite the ion (blue arrow). The decay paths are represented by the red arrows. In the experiments light from the $\pi$ decay transition (crossed in the figure) is suppressed due to the collection geometry.}
    \label{fig:YbEnergyDiagram}
\end{figure}

For our experiments we use $^{171}$Yb$^+$ because of its long-lived hyperfine ground states, relatively simple electronic structure, and fast cycling transition for  quick readout \cite{Crain_2019}. A simplified energy level structure for $^{171}$Yb$^+$ is shown is Fig.\,1 with the ion qubit states $\left|\uparrow \right>$ = $^2$S$_{1/2}$, $\left|F=1,m_F=0\right>$ and $\left|\downarrow \right>$ = $^2$S$_{1/2}$ $\left|0,0\right>$, and the photonic qubit states, $\left| \nu_0 \right>$  and $\left| \nu_1 \right>$, which are separated by 12.6 GHz.

For the ion-photon entanglement protocol \cite{Moehring_2007B} we begin by optically pumping the $^{171}$Yb$^+$ ion into $^2$S$_{1/2} \left|0,0\right>$ and then transferring population into $^2$S$_{1/2} \left|1,-1\right>$ with a microwave $\pi$-pulse at 12.6 GHz. The ion is then excited into the $^2$P$_{1/2} \left|1,-1\right>$ state via a $\pi$ transition driven by a short, linearly polarized, resonant laser pulse. From this state the ion can spontaneously decay via three different paths shown in Fig.\, 1. With the appropriate orientation of the quantization axis and fluorescence collection geometry, the spontaneously emitted photons from the $\pi$ transition are not emitted in the collection direction and hence are geometrically suppressed.

After decay, the ion-photon system is in the state $\left( \left|\uparrow, \nu_1\right>+e^{i\phi} \left|\downarrow, \nu_0\right> \right)/\sqrt{2}$ where the state of the ion is entangled with the frequency of the photon. Measurements in the logic basis show the correlation between the states of the two sub-systems, but measurement in at least one other basis is necessary to unconditionally prove entanglement. While arbitrary rotations of the ion qubit with microwaves is straightforward to realise, this is much more difficult to achieve for photonic frequency qubits as it would require high-efficiency active microwave components to change the photon's energy. Our set-up currently is not capable of implementing single qubit rotations on frequency photonic qubit and thus can only be used to verify correlations.

\section{Experimental Setup}

\begin{figure}[ht!]
    \centering
    \includegraphics{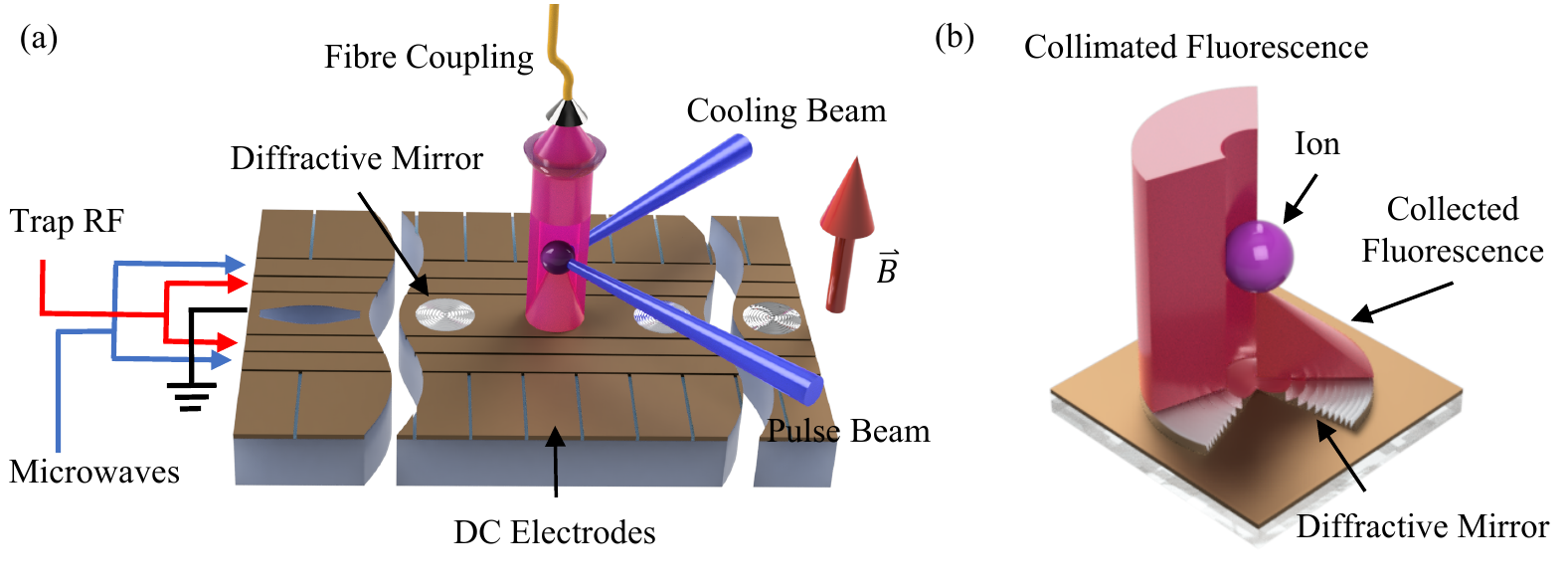}
    \caption{(a) Schematic of the trapping set-up. Cooling and pulsed excitation beams are perpendicular to each other. The DC electrodes allows shuttling of the ion along the trap. The magnetic field $\vec{B}$ is perpendicular to the trap's surface and defines the ion's quantization axis. (b) Image of the diffractive mirror on the trap's surface. A cross section is shown for the collection of emitted ion light and for the resulting collimated light.}
    \label{fig:ChipTrapExptConfig}
\end{figure}

Individual $^{171}$Yb$^+$ ions are confined above a micro-fabricated surface trap \cite{Shappert_2013} shown in Fig.\,2a. The trap design includes several integrated diffractive mirrors for collection of the ion fluorescence and coupling the light into a single mode optical fibre. Ions are produced from a neutral Yb beam via isotope selective two-photon photoionization \cite{Blums_2020}. They are trapped at the electric field node generated 60 $\mu$m above the trap surface from two RF rails and shuttled along the trap by varying the voltages on 48 segmented DC electrodes (see Fig.\,2a). The DC electrodes are independently controlled and are used to shuttle the ion over the top of the different diffractive mirrors.

Fig.\,2a shows the geometry of the ion trap, together with the cooling beam, pulsed excitation beam, magnetic field used to define the quantization axis, and the photon collection path. Doppler cooling is performed by having the cooling laser light nearly resonant with the $^2$S$_{1/2}$ F = 1 $\leftrightarrow$ $^2$P$_{1/2}$ F=0 transition with a 14.7 GHz repumping sideband generated by an EOM (Qubig: PM-Yb+$\_$14.7) to clear off resonant scattering into $^2$S$_{1/2}$ F = 0, the light has a linear polarization roughly 54.7$^{\circ}$ from the quantization axis to prevent unwanted coherent population trapping during cooling \cite{Ejtemaee_2010}. Omitted for clarity is a 935 nm repumping laser co-linear with the cooling beam to clear the meta-stable $^2$D$_{3/2}$ states and the 399 nm photoionization laser.

The diffractive mirrors etched into the central ground rail have dimensions of $80 \mu$m $\times$ 127 $\mu$m, covering 13.3\% of the total fluorescence solid angle. The mirror used for our experiment has a focal length of 60 $\mu$m (corresponding to an average NA of 0.68), generating a collimated beam from the ion’s fluorescence (see Fig.\,2b). The fluorescence passes through $\lambda$/4 and $\lambda$/2 waveplates angled so the $\sigma^-$ photonic qubits are transmitted through a polarizing beamsplitter. During cooling and state readout fluorescence from $\sigma^+$ transition decays on $^2$P$_{1/2} \left| F=0, m_F = 0 \right>$\,$\rightarrow$\,$^2$S$_{1/2} \left| F = 1 , m_F =-1\right>$ is reflected by this beamsplitter and is directed onto an avalanche photodiode (APD) (Laser Components: COUNT-100B) for detection. The transmitted photonic qubit fluorescence is coupled into a single mode polarization maintaining fibre (Thorlabs: PM-S350-HP) via a mode-matching telescope reaching an overall ion to fibre coupling efficiency of 2.7(3)\%.

From the fibre, the photon frequency is then measured using the UV hyperfine spectrometer (Fig.\,3). After the fibre a f = 4.02 mm aspheric lens (Thorlabs: C671TME) collimates the beam and sends it to an expanding telescope made of a second f = 4.02\,mm asphere and a 75 mm diameter, f = 500 mm plano-convex lens (Thorlabs: LA4246-UV). After these optics, light is sent to the key component of this apparatus, a 128 mm $\times$ 102 mm Richardson holographic grating with 4320 lines/mm. The grating is operated in a near Littrow configuration at an angle of 59$^{\circ}$. The reflected beam is slightly offset from the input path so it can be picked off using a D shaped mirror and sent along the detection path. A flip mirror is used to send the ion light to either a camera set-up, for beam profiling and diagnostics, or a mirrored knife edge type prism (ThorLabs: MRAK25-F01) to separate the two qubit frequencies for detection. After the mirrored prism, the two frequency beams are focused through 150 $\mu$m pinholes and detected on photomultiplier tubes (PMT) (Hamamatsu: H10682-210) which are otherwise protected from background light by 370 nm line filters (Semrock Brightline: FF01-370/10-25). Pulses from both detectors are counted using a time to digital converter (ID Quantique: id800-TDC).

For the experiment we adjusted the position of the first 4.02 mm asphere so that a 22(1) mm diameter beam reaches the 500 mm lens. This third lens is then moved in order to produce a collimated beam that is then reflected by the grating. This configuration has a resolving power of around $\mathcal{R} =\Delta \nu/\nu\sim$95,000, with roughly that many lines of the grating illuminated.

\begin{figure}[ht!]
    \centering
    \includegraphics{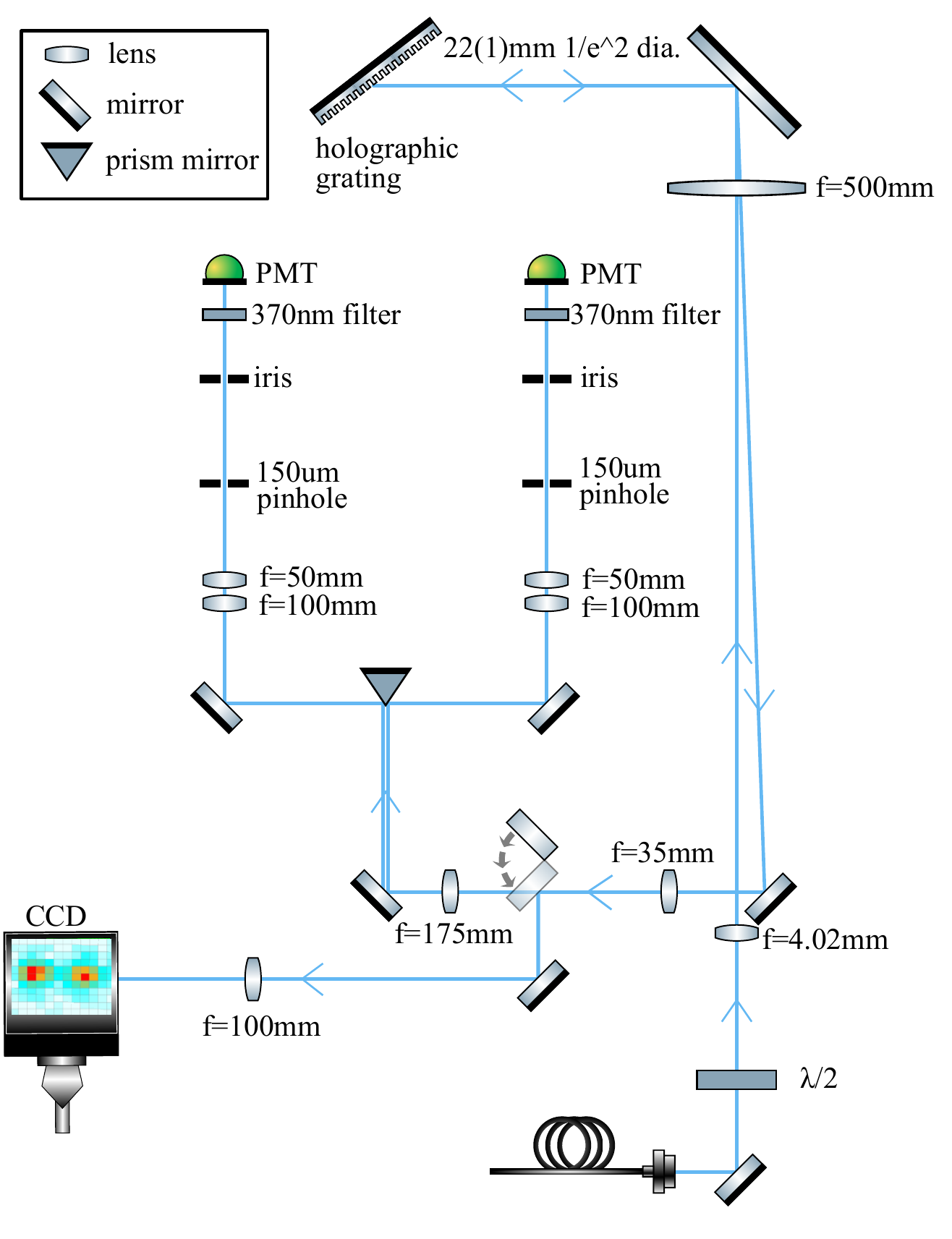}
    \caption{Diffraction grating spectrometer set-up. Ion light is inserted in the bottom of the image and is collimated with an asphere. Light is expanded with a telescope and hits the holographic grating in Littrow configuration where it is reflected on a slightly different trajectory. This light is picked off for detection, either to a CCD or to a mirrored prism which separates the two frequencies of light to be incident on two PMTs.}
    \label{fig:GratingSpectrometer}
\end{figure}

We characterised the spectrometer by measuring its losses and the spot separation between two laser frequencies corresponding to $\left|\nu_1\right>$ and $\left|\nu_0\right>$, centred around 369.5 nm (811.3 THz), and separated by 12.6 GHz. Fig.\,4 shows the two spots measured by an EMCCD camera (Andor: Luca-S 658M), the spot $1/e^2$ diameters, 47.4(4) $\mu$m and 46.6(4) $\mu$m, and a spot separation of 3.5(2) radii. Theoretically, the field-mode overlap of these two Gaussian beams is $<$0.34\% corresponding to a maximum estimated fidelity of $>$99.66\%, but sending attenuated laser light to the two PMTs, we measured a detection fidelity of 98.0(6)\% for $\left|\nu_0\right>$ and 97.2(4)\% for $\left|\nu_1\right>$. The spectrometer system's overall quantum efficiency, from fibre to detector, is 3.7(5)\%. This number accounts for 79(2)\% coupling efficiency into a single mode optical fibre, 25(3)\% grating efficiency and losses in the set-up optics, and 19\% PMT quantum efficiency.
 
\begin{figure}
    \centering
    \includegraphics{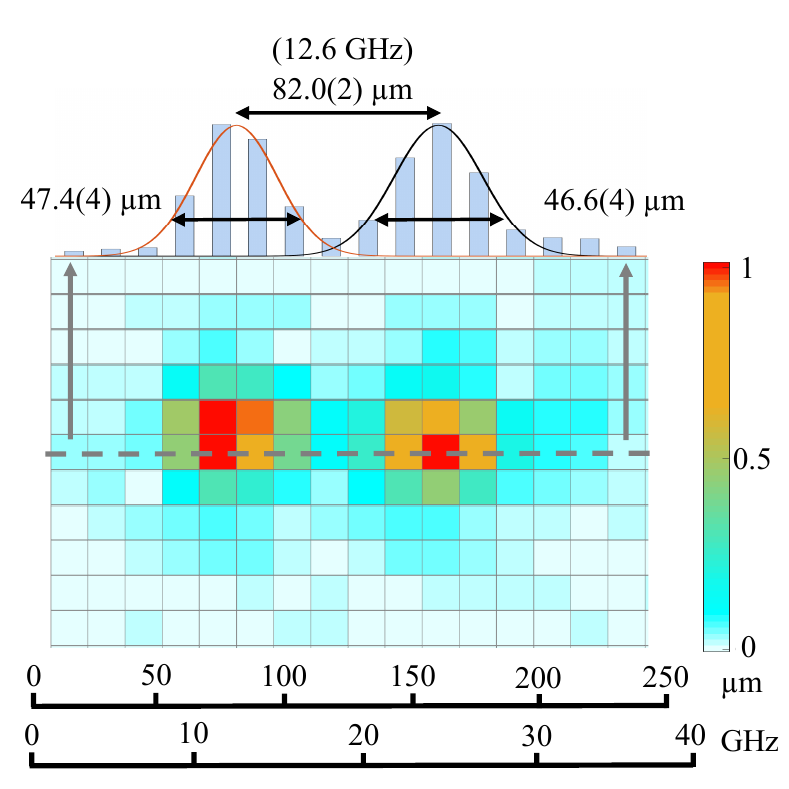}
    \caption{Spectrometer spot separation of laser photons with 12.6 GHz frequency difference. The spots are 47.4(4) $\mu$m and 46.6(4) $\mu$m $1/e^2$ diameter with a separation of 82.0(2) $\mu$m, corresponding to 3.5(2) $1/e^2$ radii and a resolution of 3.6(2) GHz.  The image was taken with a 1 sec exposure.}
    \label{fig:HFSplitSpots}
\end{figure}

\section{Results}

For our experiment we followed the protocol described in Sec. 2 and ref. \cite{Moehring_2007}. Initially the ion is prepared in state $\left| \downarrow\right>$ via optical pumping for 10 $\mu$s by the cooling laser, modulated by a 2.1 GHz EOM (Qubig: PM-Yb+$\_$2.1), resonant with the $^2$S$_{1/2}$ $\left|1,-1\;0\:+1\right>$ $\leftrightarrow$ $^2$P$_{1/2}$ $\left|0,0\right>$ and $^2$P$_{1/2} \left|1,-1\;0\:+1\right>$ transitions. Subsequently a 17 $\mu$s microwave $\pi$-pulse at 12.637855 GHz transfers the population from $\left|\downarrow\right>$ to $^2$S$_{1/2} \left|1,-1\right>$ with 91(4)\% fidelity. The lower than ideal fidelity is due to a 9(2) kHz detuning of the microwave from resonance.

The ion is then optically excited from $^2$S$_{1/2} \left|1,-1\right>$ to $^2$P$_{1/2} \left|1, -1\right>$ by a linearly polarized pulse (see Fig.\,1 and Fig.\,2a). The pulse is created from a CW laser by an AOM (AAoptics MQ180-A0,2-UV) which generates a 51 ns optical pulse. From the state $^2$P$_{1/2} \left|1, -1\right>$, the ion decays to the ground state with a spontaneous lifetime of 8.1 ns and emits a single photon at 369.5 nm. Using a Lindbladian describing the system in Fig.\,1 we find the probability the ion excites and decays into $\left| \uparrow , \nu_1 \right>$ or $\left| \downarrow , \nu_0 \right>$ after a pulse is estimated as 11.6(4)\% for either state.

The ion fluorescence is collimated by the diffractive mirror on the chip trap surface, covering a solid angle of 14.3\% on the $\sigma$ transition, and collected in a direction parallel to the magnetic field axis (see Fig.\,2a and 2b). This configuration minimizes the light collected from the $\pi$ decay channel into $^2$S$_{1/2} \left|1,-1\right>$, which is also partially filtered out by a polarizer, and optimizes the collection of light from the $\sigma^-$ decay into $\left| \uparrow \right>$ and $\left| \downarrow \right>$. As a result of this decay an entangled state is generated between the ion and the photon. The emitted circularly polarized single photons are rotated to linear polarization with a quarter wave plate, sent through a polarizing beam splitter cube to spatially separate them from photons with a different polarization, and coupled into a single mode fibre. The fibre is connected to the UV hyperfine spectrometer and the photon frequency states, $\left|\nu_1 \right>$ and $\left|\nu_0 \right>$, are spatially separated to be read out on separate PMTs (see Fig.\,3) with a 200 ns time gate.

The ion state is read out by shining the cooling laser without any sidebands from the EOMs, for 1.38 ms and measuring the arrival times of fluorescence counts on a third "readout" APD. Using the state selective detection method outlined in \cite{W_lk_2015}, the trapped ion qubit state was determined with an approximate readout fidelity of 95.5\% for $\left|\uparrow\right>$ and 97.3\% for $\left|\downarrow\right>$. 

We ran this protocol over 5 hrs and 50 minutes in 10 min blocks with each run taking 1411 $\mu$s. From these runs we detected a total of 2006 signal photons on the spectrometer. For each successful click from the signal photons, the ion qubit state was also recorded and the ion-photon quibit correlation fidelity is calculated. The coincidence detection probability per shot for our system is 0.013(3)\%, measured over 14,883,327 runs. This probability comes from a 91(4)\% state preparation fidelity, the 11.6(4)\% probability of generating a photon from the excitation pulse, a 2.7(3) \% probability of collecting the photon into the fibre and the 3.7(5)\% efficiency of the spectrometer set-up. Combined, the expected coincidence detection probability per shot is in agreement at 0.011(2)\%.
	
\begin{figure}
    \centering
    \includegraphics{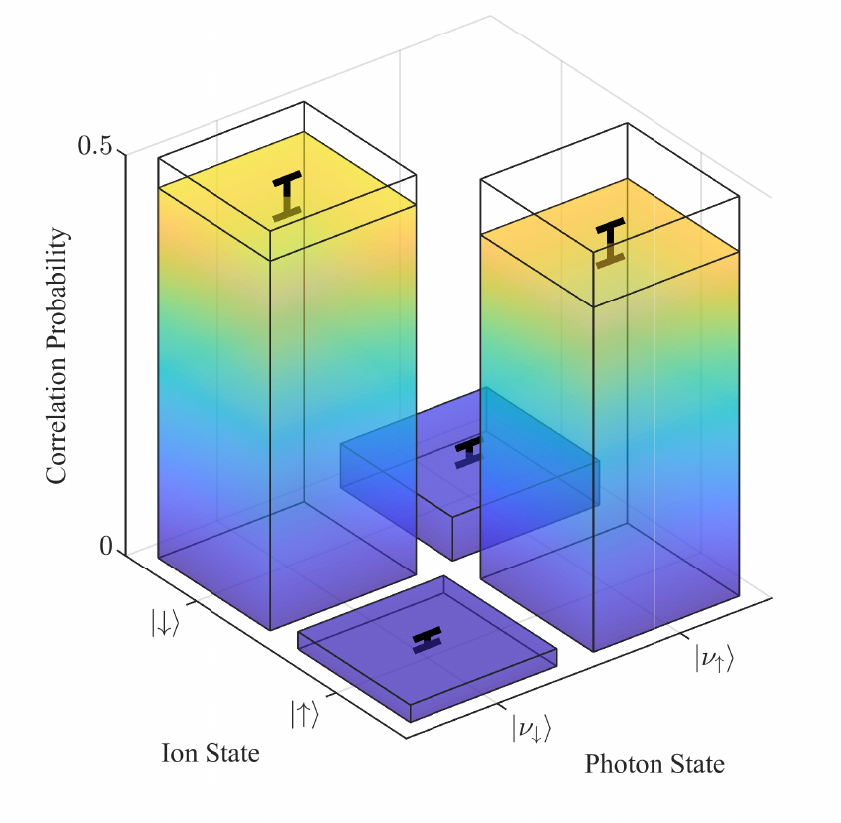}
    \caption{Ion photon correlations. Wire frames indicate ideal outcome. For a $\left|\nu_0\right>$ photon we get 95.2(7)\%  and for a $\left|\nu_1\right>$  photon we get 89.6(10)\% fidelity. Data corresponds to 2006 counts out of 14,883,327 experimental runs.}
    \label{fig:StateCorrelations}
\end{figure}	 

The correlations between the states of the ion and the signal photon are shown in Fig.\,5. For the $\left|\downarrow, \nu_0\right>$ state fidelity is 95.2(7)\% and $\left|\uparrow, \nu_1\right>$ state fidelity is 89.6(10)\% while the average fidelity of the correlation matrix is 92.4(8)\%. The asymmetry in fidelity is mainly due to higher rate of incorrect photon readout of the $\left|\nu_1\right>$ photon. The main sources of infidelity are trapped ion qubit hyperfine state readout at approximately 3.6\%, and spot separation in the diffraction grating spectrometer at 2.4(5)\% (see Fig.\,4).

\section{Conclusions}

We demonstrate ion-photonic frequency qubit correlations with a 92.4(8)\% average fidelity. Measurements were performed with a high-resolution spectrometer for characterizing frequency encoded photonic qubits. This is an important result for the development of an ion-based quantum network which use more robust frequency encoding. The next step would be to demonstrate quantum frequency conversion of our photonic qubit and show that correlations are preserved after frequency conversion. These are key first steps towards developing quantum repeater architecture for photonic qubits in a fibre optic network.

Several improvements could be made to the current system for improving its efficiency and fidelity. State preparation and optical pulse fidelity can be improved to near unity by shaping our microwave pulse (by using a BB1 protocol \cite{Shappert_2013} for example) and using an EOM for switching the optical pulse duration to below the excited lifetime of 8.1 ns. Readout can be improved to 99.4\% fidelity with lower dark count rate detectors and readout can be performed in 176 $\mu$s \cite{Crain_2019} increasing our experimental cycle speed to ~5 kHz. Once we have our single photon generation protocol optimized, we can send ion fluorescence directly to the grating, negating the fibre coupling and attenuation losses. The grating performance can be improved with larger optics which will allow us to get a diffraction limited spot without loss of efficiency, so the grating efficiency will be the ideal diffraction efficiency of 55\% with correct polarization and the fidelity will be $<$99.9\%. By making these changes we expect to achieve a fidelity of 99.1\% and achieve a bit rate of ~34 Hz for successful trials without considering detector quantum efficiency.

\section{Acknowledgements}
This research was financially supported by the Griffith University Research Infrastructure Program, the Griffith Sciences equipment scheme, Australian Research Council Linkage Project (LP180100096). ML was supported by Australian Research Council Future Fellowships (FT180100055), VB, JS, KS, and SC were supported by the Australian Government Research Training Program Scholarship.

\section{Bibliography}

\bibliographystyle{iopart-num.bst}
\bibliography{GratingPaperRefs}

\end{document}